# Bone marrow sparing for cervical cancer radiotherapy on multimodality medical images

Yuening Wang, Ying Sun, Jie Yuan, Kexin Gan, Hanzi Xu, Han Gao, Xiuming Zhang

*Abstract*—**Cervical cancer threatens the health of women seriously. Radiotherapy is one of the main therapy methods but with high risk of acute hematologic toxicity. Delineating the bone marrow (BM) for sparing using computer tomography (CT) images to plan before radiotherapy can effectively avoid this risk. Comparing with magnetic resonance (MR) images, CT lacks the ability to express the activity of BM. Thus, in current clinical practice, medical practitioners manually delineate the BM on CT images by corresponding to MR images. However, the time-consuming delineating BM by hand cannot guarantee the accuracy due to the inconsistency of the CT-MR multimodal images. In this study, we propose a multimodal image oriented automatic registration method for pelvic BM sparing, which consists of three-dimensional bone point cloud reconstruction, a local spherical system iteration closest point registration for marking BM on CT images. Experiments on patient dataset reveal that our proposed method can enhance the multimodal image registration accuracy and efficiency for medical practitioners in sparing BM of cervical cancer radiotherapy. The method proposed in this contribution might also provide references for similar studies in other clinical application.**

## I. INTRODUCTION

THE number of women who are diagnosed with cervical cancer is more than half a million each year [1]. Cervical cancer threatens female health and the disease brings about over 300,000 deaths globally [2]. There are growing numbers of people with cervical cancer treated with three-dimensional (3D) conformal radiotherapy and intensity-modulated radiotherapy [3]. Pelvic radiotherapy has been regards as standard therapy for advanced stage disease of cervical cancer, which greatly improve the overall five-year survival rate after therapy [4]. However, the does delivered to cure severe tumors might exceed the constrain of toxicity in human body. Thus, the medical practitioners need to control the dose intensity of radiotherapy. Since approximately 40% of the human body bone marrow resides within the pelvic bones which are irradiated in the range of pelvic radiotherapy [3], high rates of severe late complications of the therapy may occur [5]. In order to provide living quality after treatment, efforts to improve normal tissue sparing include the utilization of distinct radiation-fractionation schedules and intensity-modulated radiotherapy, which can cut down the radiation damage of bone marrow in contrast to conventional radiotherapy.

In clinical practice, Water-fat magnetic resonance imaging (MRI) is sensitive to changes in bone marrow composition [6], medical practitioner can recognize the BM through corresponding Water-fat MR images [7]. Hence it has the ability to provide contrast to distinguish BM with different levels of activity based on various fat content in BM. However, the images used for radiotherapy planning does come from computed tomography (CT), which is determined by Treatment Planning System [8]. By delineating BM in CT images with corresponding Water-fat MR images, the radiation dose delivered in a single treatment volume of a process can be correctly evaluated, which can reduce the irradiated damage of BM [7].

Distinguishing structure similarity difference between CT-MR multimodal images and matching image sequence are time-consuming and labor-intensive, which challenges the medical practitioners when labeling BM. As shown in Fig.1(b) and Fig.1(c), the versatile expression pattern of the same tissue in multimodal images hampers the progress of the BM delineation by hand. Furthermore, CT images and Water-fat MR images are both two-dimensional sequences, the image range and the slice interval of the same patient are different which causes the mapping difficulty when matching multimodal images.

Medical image registration is of great importance for clinical diagnosis and therapy. Image-guided intervention mainly depends on image registration [9]. Telesurgery, Image-Guided Radiotherapy (IGRT), and precision medicine all rely on accurate image registration. Medical studies also involve image registration, such as motion tracking, segmentation, dose accumulation, image reconstruction [10]. Multimodal image fusion is the principal method of registration in medical images which overlays the information among different imaging modes. The difficulties of multimodal medical image registration

Yuening Wang and Ying Sun contribute equally in this study. Corresponding author are Jie Yuan and Hanzi Xu (Email: yuanjie@nju.edu.cn and xuhanzi@njmu.edu.cn).

Jie Yuan, Yuening Wang, Ying Sun and Kexin Gan are with the School of Electronic Science and Engineering, Nanjing University, Nanjing 210023, China. (Email: yuanjie@nju.edu.cn, wangyuening@smail.nju.edu.cn,sunying@smail.nju.edu.cn,181180029 @smail.nju.edu.cn). Hanzi Xu, Han Gao, Xiuming Zhang are with The Jiangsu Cancer Hospital, Nanjing, China. (Email: xuhanzi@njmu.edu.cn, 312572057@qq.com, zhangxiuming360@163.com)

include building the suitable image similarity measure index because of appearance diversities between different imaging modalities, the image variation for the same person using the same imaging method for the reason of metabolic processes at different time [11].

Multimodal medical image registration is the procedure of aligning two or more modalities of imaging in the same geometrical coordination system [10]. The purpose of the process is to find an optimum spatial transformation that can be employed to match the structures-of-interest of different images in the best way. Medical image registration has a wide application in almost all parts of human body [12]. Multimodal image registration methods can be classified into area-based methods, feature-based methods and learning-based methods [13]. In the area-based methods, the intensity information of multimodal medical image is the key to match images, so the similarity measure assessment has the great influence on the registration. There are two common measure assessment metrics for multimodal medical image registration, modality-independent measure metric and modalities normalization-based metric. Mutual information (MI) is the basic concept from information theory, which has the advantage in multimodal image registration. Maes present the method that uses the maximization of MI as the indicators of geometrically aligned images and MI is applied to measure the statistical dependence or information redundancy in the multimodal image intensities [14]. However, it's the challenge to keep the robustness in the process of matching images owing to different intensity of the same structure in multimodal images. Featured-based methods contains the procedure of feature detection, feature description and feature matching. Point feature has a wide application in multimodal image registration compared to line and region feature. Point set registration (PSR) iteratively estimates the transformation matrix based on a global transformation model between fixed point sets and moving point sets. Iterative closest point (ICP) is one of the most representative algorithms [15]. ICP algorithm concentrates on Euclidean distance to optimize the transformation matrix. Chen improved the original ICP with point-to-plane distance, which relies on the distance from points to planes to optimize registration and does not require point-to-point correspondence [16]. Segal proposed a Generalized-ICP (GICP) algorithm, which introduced a probabilistic model combined with ICP and point-to-plane ICP, which maintains the simplicity of ICP algorithm and reduce uncertainty of registration associated with parameter selection [17]. For the learning-based methods, Herring presented an image registration method based on convolutional neural network (CNN) to predict the spatial deformation that matches moving image to fixed image [18]. Hu exploited a GAN-based model relying on segmented label similarity and the adversarial loss to register multimodal medical images [19]. The learning-based registration technologies is promising with the development of deep-learning. However, severely lack of sufficient dataset limits the learning-based registration methods [13].

In this study, we propose a method in BM sparing radiotherapy for cervical cancer involving Pelvic bone registration between MR and CT sequential images. Bone regions are extracted from multimodal image sequences to construct three-dimensional (3D) bone point cloud in based on semantic segmentation network. Slices-to-volume image registration are performed consequently for BM delineation with a proposed algorithm which adopts joint coordinate system and neighborhood information. When applied in comparison experiments with other algorithms, our proposed method can yield reliable and better registration accuracy in both 3D point cloud and image sequences, while avoiding the complex similarity measurement function construction and feature points selection.

## II. METHODS

### A. Data Acquisition

In this work, three kinds of medical images are involved, which are supported by the Department of Radiation Oncology and Department of Radiology, Jiangsu Cancer Hospital. All the experimental protocols of this work are implemented under the approval and guidance of Jiangsu Cancer Hospital, and the relevant clinical norms and medical ethics were strictly followed to ensure that patients' privacy was not violated, disclosed, and patients' rights and interests were not harmed. The CT images of pelvic cavity are obtained from GE's discovery CT750HDCT or Hispeed NX/i scanner, and the tube voltage was 120kV. The MR images of pelvic cavity are obtained from Philips Ingenia 3.0 T with scanning sequence of T1WI. The fat fraction (FF) images of pelvic cavity are collected from Philips Ingenia 3.0 T with fat-water separated Dixon method.

The human pelvic cavity bone of interest in the study is shown in Fig.1(a), which includes sacrum, ilium and coccyx. Fig.1(b) shows the CT image in grayscale, in which the structure of bones is highlighted in white. Fig.1(c) and Fig.1(d) respectively show the MR image in grayscale and Water-fat MR image in RGB-channel figure. The MR image contains abundant anatomical structure and tissue information, in which the clear outline is usually surrounded around the bone. The color bar in the Water-fat MR image represents the fat density

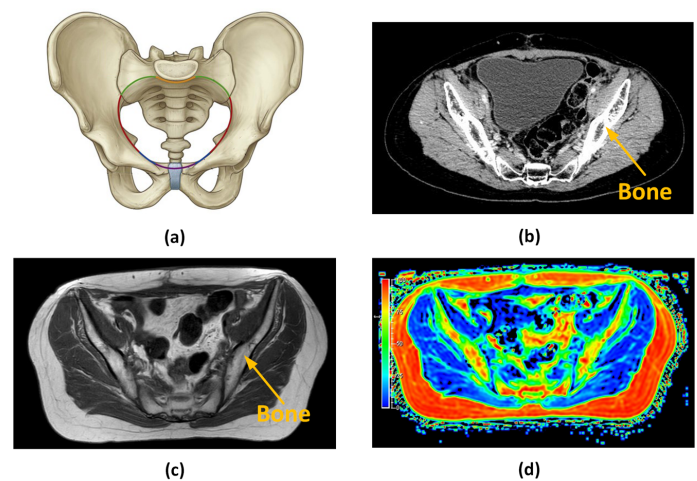

Fig.1 (a) Anatomical drawing of human pelvic cavity bones. (b) The CT image of pelvic cavity. (c) The MR image of pelvic cavity. (d) The fat fraction image of pelvic cavity calculated from Water-fat MR image.

content. The color closer to red denotes the higher fat density, and the color closer to blue denotes the lower fat density, which is the important basis of BM sparing for medical practitioners. Although the image of MR and Water-fat MR are different, they share the same structure in bone area due to the same source. In this work, the bone of pelvic cavity constructed from CT and MR image sequences are the target for registration.

### B. Construction of bone point cloud

Automated image registration can help the medical practitioners in sparing pelvic bone marrow of cervical cancer radiotherapy. Water-fat MR images can help medical practitioners mark the bone marrow according to the fat density information. However, it's difficult to identify the structure of bone from Water-fat MR images. In this work, the bone regions are segmented respectively from CT images and MR images which shared the same structure with Water-fat MR images. Due to different imaging methods, tissues show different appearances of in CT and MR images, but the bone is easier to identify in both CT and MR images, which is regarded as rigid body.

As shown in Fig.2, the process of construction of bone point cloud from MR images and CT images contains three stages: bone region extraction, interpolation between sequence images and generation of bone point cloud. The bone regions are characterized with a highlighted white in CT images and the bone regions in MR images are usually recognized with clear outlines. In our work, the ground truth used for the semantic segmentation training network is manually labeled by experienced radiologists. The labeling task is completed using LabelMe [20]. The CT images have a 512*512 resolution while the resolution of MR images is uniformed to 512*512 resolution from 784*784, 704*704 and 576*576.

The semantic segmentation network applied to the extraction of bones is the U-Net3+ network [21], which is based on the U-Net network [22]. The framework of U-Net network includes three main components: the main feature extraction network, the intensive feature extraction network and predicting network. Based on the Encoder-Decoder Structure and skip connection, U-Net can have the full use of low-level and high-level semantic features. U-Net3+ is improved on the basis of U-Net and takes advantage of full-scale skip connections and deep supervisions, which has the improvement on segmentation for our dataset compared to U-Net. Due to the input images with structural similarity difference between CT and MR images, two semantic segmentation networks are respectively trained in the CT image datasets and the MR image datasets. The predicting output images consisting of the bone regions are the original information for building point cloud. For the simplification of description, the predicting output images with bone regions are called bone images in this work. The extraction of bone regions with the segmentation network can reduce the complexity of registration and improve the registration accuracy, which avoiding the direct registration of the multimodal images of the whole pelvic.

The 2D sequential bone images obtained by segmentation network prediction need preprocessing. As shown in Fig.2, the X-Y plane is denoted as the transverse section and the Z-axis direction is denoted as the vertical direction to the transverse section. The X-Y plane scale of 2D sequential bone images is zoomed by the maximum bone length in Y-axis direction. The scale of MR bone images is fixed, and the scale of CT bone images is changed according to the proportional relationship between the maximum bone lengths in Y-axis direction of CT and MR bone images. The Z-axis resolution of generated point cloud relies on the slice spacing and the X-Y plane resolution of generated point cloud is the same as the resolution of the corresponding image sequences, which will lead to the scale deformation in contrast to the actual human pelvic cavity bone. In this work, the cubic interpolation is used for eliminating resolution distortion in different directions. With the preprocess of slice interpolation, the sequential bone images are employed for the generation of bone point cloud.

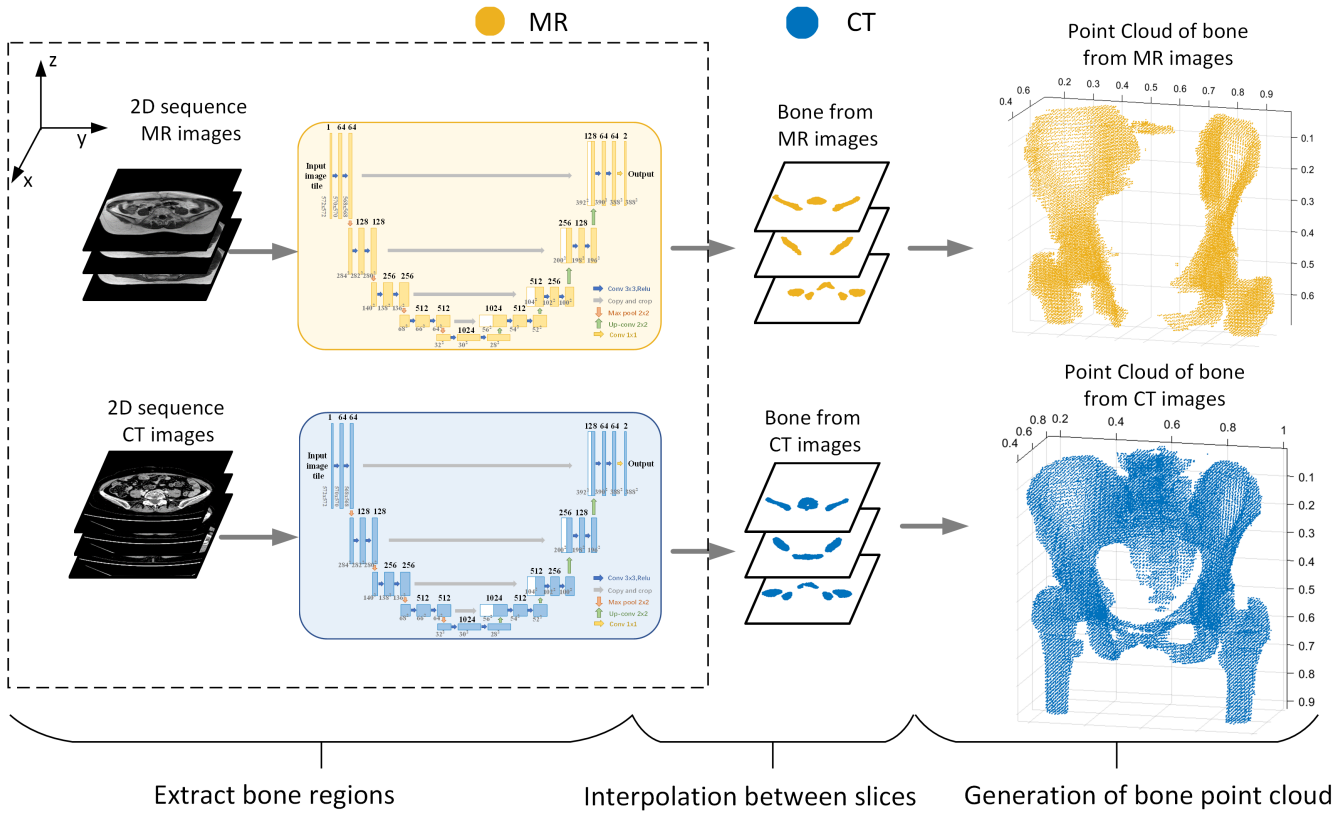

Fig.2 The process of construction of bone point cloud based on 2D sequence MR images and 2D sequence CT images. The bone point clouds of MR and CT are displayed in normalized coordinate system.

*C. Multimodal image registration*

As one of the most common point cloud registration algorithms, ICP algorithm registers fixed point cloud and moving point cloud by finding the transformation matrix that that maximize the error calculated by Euclidean distance [23]. The ICP algorithm has the advantage of easy realization and can usually obtain an accurate registration. However, the performance of the ICP algorithm depends on initial conditions of fixed-point cloud and moving-point cloud, which is possible to fall into the problem of convergence to local optimal [24]. In the conventional ICP algorithm, the error function used in the iterative optimization of the transformation matrix is only related to the distance in Cartesian coordinate system [25]. In this work, we propose a registration algorithm based on the joint coordinate system that combined global **C**artesian coordinate system with local point cloud **S**pherical coordinate system, which also contains **N**eighborhood features and partition registration related to bone characteristic, which is called CSN-ICP in abbreviation.

$P_t = \{p_{t_i}\}_{t_i=1}^{N_{P_t}}$ is the target bone cloud point set constructed from the CT sequential images, which is the fixed-point cloud in the iterations of registration. $p_{t_i}$ denotes the $t_i$th point in the target point set $P_t$, while $N_{P_t}$ denotes the total number of points in the target point set $P_t$. $P_s = \{p_{s_i}\}_{s_i=1}^{N_{P_s}}$ is the source bone point cloud constructed from the MR sequential images, which is the moving-point cloud in the iterations of registration. $p_{s_i}$ denotes the $s_i$th point in the source point set $P_s$, while $N_{P_s}$ denotes the number of points in the source point set $P_s$.

In our proposed CSN-ICP registration algorithm, for each point $p_{s_i}$ in source point set $P_s$, the point $p_{t_i}$ with the highest similarity in the target point set $P_t$ is queried as the corresponding point $p_{c_i}$ and the corresponding point set is marked as $P_c = \{p_{c_i}\}_{c_i=1}^{N_{P_c}}$. The transformation matrix that includes the rotation matrix $R$ and translation vector $T$ is updated according to the error function $E$ in the Eq.1.

$$R^*, T^* = arg\,min \frac{1}{N_{P_s}} \sum_{s_i=1}^{N_{P_s}} E(P_c, (R \cdot P_s + T)) \quad (1)$$

where $R^*$ and $T^*$ are the most optimal rotation matrix and translation vector, which satisfies the minimum error function. In conventional ICP registration algorithm, the error function is constructed in Cartesian coordinate and use the square sum of Euclidean distances between the corresponding point set and the source point set. The conventional ICP registration algorithm fails to consider the elimination of the matching points with large errors, which is sensitive to noise and lack of robustness [26].

In the conventional ICP registration, each point is described by the position in Cartesian coordinate system as $(x, y, z)$. In our proposed CSN-ICP registration algorithm, a local point cloud spherical coordinate system is introduced to describe the local features, which is constructed on the basis of the normal vector and curvature of point cloud as shown in Fig.3(a). The description of each point is extended with the local features and neighborhood information.

The input point cloud set is supposed as $P = \{p_i\}_{i=1}^{N}$. For each point $p_i$, the k-nearest neighbor point set is marked as $P_{N_i} = \{p_{N_{ij}}\}_{j=1}^{k}$. Based on the error in the least square calculation, the best fitting local plane of $P_{N_i}$ is defined as follows,

$$Plane_i(\boldsymbol{n}, d) = arg\,min \sum_{j=1}^{k} (\boldsymbol{n} \cdot p_{N_{ij}} - d)^2 \quad (2)$$

where $\boldsymbol{n}$ is the normal vector of the best fitting local plane and $d$ is the distance between origin of coordinate and the best fitting local plane [27]. The center of mass $C_i$ of the k-nearest neighbor point set $P_{N_i}$ is calculated as follows,

$$C_i = \frac{1}{k} \sum_{j=1}^{k} p_{N_{ij}} \quad (3)$$

The covariance matrix $M$ is calculated as follows,

$$M = \begin{bmatrix} p_{N_{i1}} - C_i \\ p_{N_{i2}} - C_i \\ ... \\ p_{N_{ik}} - C_i \end{bmatrix}^T \begin{bmatrix} p_{N_{i1}} - C_i \\ p_{N_{i2}} - C_i \\ ... \\ p_{N_{ik}} - C_i \end{bmatrix} \quad (4)$$

Employing eigen decomposition of the covariance matrix $M$, three eigenvalues $\lambda_1, \lambda_2$ and $\lambda_3$ can be obtained. Supposing that $\lambda_{min}$ is the smallest value of these three eigenvalues, the normal vector $\boldsymbol{n_i}$ of the point $p_i$ is the eigenvector corresponding to the smallest eigenvalue. The normal vector $\boldsymbol{n_i}$ of the point $p_i$ is described as $(n_x, n_y, n_z)$, where $n_x, n_y$ and $n_z$ are separately represented as the length of the normalized normal vector $\boldsymbol{n_i}$ on the X, Y and Z axes respectively. The curvature $\delta_i$ of the point $p_i$ is calculated as follows,

$$\delta_i = \frac{\lambda_{min}}{\lambda_1 + \lambda_2 + \lambda_3} \quad (5)$$

The origin of local point cloud spherical coordinate system is established by the position of the point $p_i$ in conventional

Cartesian coordinate system. As shown in the Fig.3(a), each point in local point cloud spherical coordinate system can be described as the distance $r_i$, elevation angle $\theta_i$ and the azimuth angle $\varphi_i$.

$$r_i = \delta_i$$
$$\varphi = \arctan\frac{n_y}{n_x}$$
$$\theta = \arccos n_z \quad (6)$$

The distance $d_s$ between two points $p_1(x_1, y_1, z_1, r_1, \theta_1, \varphi_1)$ and $p_2(x_2, y_2, z_2, r_2, \theta_2, \varphi_2)$ in the local point cloud spherical coordinate system can be defined as follows,

$$d_s = abs(r_1 - r_2) + abs(\varphi_1 - \varphi_2) + abs(\theta_1 - \theta_2) \quad (7)$$

The distance $d_c$ is defined as Euclidean distance in the global Cartesian coordinate, which is defined as follows,

$$d_c = \sqrt{(x_1 - x_2)^2 + (y_1 - y_2)^2 + (z_1 - z_2)^2} \quad (8)$$

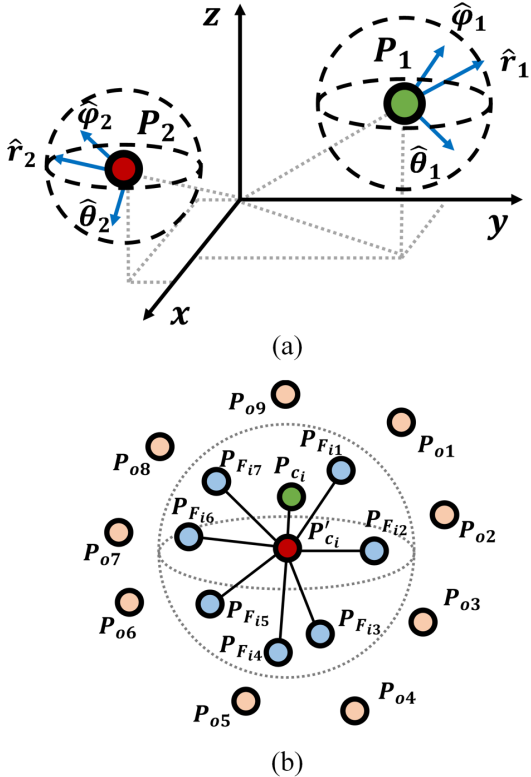

(a)

(b)

Fig.3 (a) The demonstration of the joint coordinate system that combined global Cartesian coordinate system with local point cloud spherical coordinate system in CSN-ICP registration algorithm. (b) The demonstration of neighborhood features constructed by neighbor point set with the threshold radius of neighborhood.

Each cloud point is described as $(x, y, z, r, \varphi, \theta)$ in the joint coordinate system that combined global Cartesian coordinate system with local point cloud spherical coordinate system. In our proposed CSN-ICP registration algorithm, the primary corresponding point set $P'_c = \{p'_{c_i}\}_{c_i=1}^{N_{P_c}}$ is calculated with the minimum Euclidean distance in the global Cartesian coordinate system and each point $p_{s_i}$ in the source point set $P_s = \{p_{s_i}\}_{s_i=1}^{N_{P_s}}$ can search the primary corresponding point $p'_{c_i}$ in the target point set $P_t$. Then the neighborhood point set $P_{F_i} = \{p_{F_{ij}}\}_{j=1}^{N_r}$ includes the points in the sphere with $r_{th}$ as radius and $p'_{c_i}$ as the center of the sphere, which is applied to define neighborhood features as shown in Fig.3(b). The point $p_{c_i}$ is searched based on the minimum distance $d_s$ with the point $p_{s_i}$ among the neighborhood point set $P_{F_i}$.

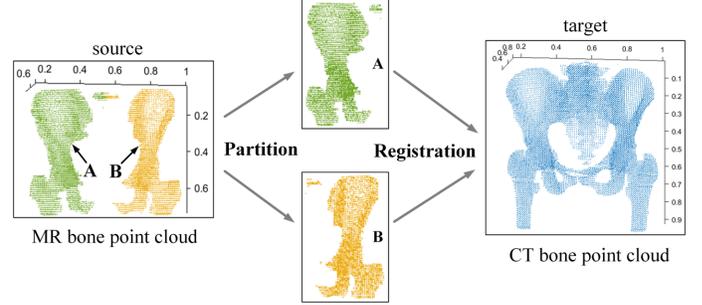

Fig.4 The schematic diagram of the division of the source point cloud in the partition registration of CSN-ICP registration algorithm. The bone point cloud is displayed in normalized coordinate system.

For each point $p'_{c_i}$ in the primary corresponding point set $P'_c$, the point $p_{c_i}$ instead of the point $p'_{c_i}$ is regards as the corresponding point. $P_c = \{p_{c_i}\}_{c_i=1}^{N_{P_c}}$ is the corresponding point set which is obtained based on the processing mentioned above. Then the corresponding point pair in the source point set $P_s$ and the corresponding point set $P_c$ can be removed by setting thresholds of the distance $d_s$ and $d_c$. Through singular value decomposition (SVD), the rotation matrix and translation vector can be acquired.

The partition registration is considered due to the motion of pelvic bone. In order to simulate the movement of pelvic bone, the source point set is divided into several parts and each source point subset registers with the target point cloud separately. Dividing the source point cloud into different parts in X-Y plane is consistent with bone expansion and displacement motion, as shown in Fig.4.

## III. EXPERIMENTS AND ANALYSIS

### A. Data Acquisition

Two semantic segmentation networks for bone regions extraction are trained: The network based on 6,361 CT images and the network based on 1,516 MR images from anonymous patients. The training dataset of semantic segmentation network is independent which means the CT images and MR images can come from different patients. The testing dataset of semantic segmentation network is also the data employed for the construction and registration of bone point cloud. For the bone registration, we include 618 CT images with slice spacing of 1.25mm, 112 CT images with slice spacing of 5mm, 240 MR images and 240 Water-fat MR images.

## B. Quality assessment

The semantic segmentation pixels can be divided into the four categories: true-positive (TP), false-positive (FP), true-negative (TN), and false-negative (FN), which form the confusion matrix [28]. The symbol abbreviation is defined in Table 1.

TABLE 1
DEFINITION OF ABBREVIATIONS OF THE CONFUSION MATRIX

| Category | Actual bone pixel | Actual non-bone pixel |
|---|---|---|
| Predicted bone pixel | True Positive (TP) | False Positive (FP) |
| Predicted non-bone pixel | False Negative (FN) | True Negative (TN) |

According to Table 1, the IOU and Dice similarity coefficient are introduced to evaluate segmentation results. They are respectively defined as follows,

$$IOU(G,P) = \frac{G \cap P}{G \cup P} = \frac{TP}{TP+FN+FP} \quad (9)$$

$$Dice(G,P) = \frac{2(G \cap P)}{G+P} = \frac{2*TP}{(TP+FN)+(TP+FP)} \quad (10)$$

where $G$ represents the ground truth of bone regions and $P$ represents the predicted bone regions.

The root mean squared error (RMSE) is introduced to evaluate point cloud registration accuracy, which is defined as follows.

$$RMSE = \sum_{i=1}^{n_{MR}} \frac{\sqrt{(x_{CTi}-x_{MRi})^2+(y_{CTi}-y_{MRi})^2+(z_{CTi}-z_{MRi})^2}}{n_{MR}} \quad (11)$$

where $n_{MR}$ represents the number of points in the MR bone point cloud. $p_{CTi}(x_{CTi}, y_{CTi}, z_{CTi})$ is the symbol of one point in the CT bone point cloud and also the target point set. $p_{MRi}(x_{MRi}, y_{MRi}, z_{MRi})$ is the symbol of one point in the MR bone point cloud and also the source point set. The point $p_{CTi}$ is the closest point on the Euclidean distance in target point set for the point $p_{MRi}$.

In this work, the point cloud registration is used for the planning CT delineation. Furthermore, the evaluation index for assessing the CT and MR bone images generated by the 3D point cloud after registration is also evaluated. Based on the quantitative description of the region beyond the common registration region, two evaluation indicators $D_{MR}$ and $D_{CT}$ are introduced in this work,

$$D_{MR} = 1 - \frac{A_{MR} \cap A_{CT}}{A_{MR}} \quad (12)$$

$$D_{CT} = 1 - \frac{A_{MR} \cap A_{CT}}{A_{CT}} \quad (13)$$

where $A_{MR}$ represents the area of MR bone region in each MR image and $A_{CT}$ represents the area of CT bone region in each CT image. $D_{MR}$ indicates the proportion of MR bone region outside the MR and CT common bone registration region. $D_{CT}$ indicates the proportion of CT bone region outside the MR and CT common bone registration region. $D_{MR}$ and $D_{CT}$ are inversely proportional to 2D multimodal bone registration accuracy.

## C. Experimental results

Two semantic segmentation U-Net3+ network models are trained separately based on CT imaged and MR images. The segmentation accuracy is tested through three-fold cross-validation. In the data for generation bone point cloud, the segmentation evaluation IOU and Dice similarity coefficient: $IOU_{MR} = 0.803$, $Dice_{MR} = 0.891$, $IOU_{CT} = 0.889$, $Dice_{CT} = 0.941$. As shown in Fig.5, the images on the first row are MR images and the images on the second row are CT images, which are collected from transverse plane of the human pelvic bone. The sequence of images is coincident with the sequence of human anatomy structure. The areas marked in green represent the bone region prediction output of semantic segmentation network models. The manually annotated labels of coccyx bone exist deviation due to the vague outline, which leads to the decreased accuracy of bone segmentation.

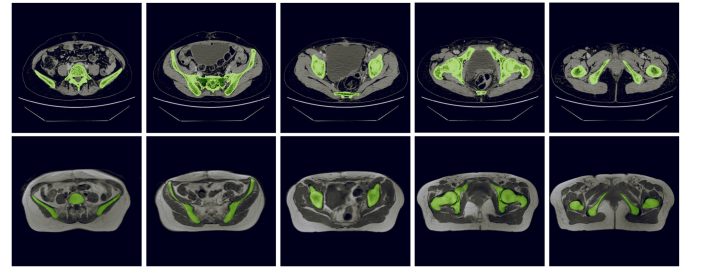

Fig.5 Prediction of bone region based on semantic segmentation U-Net3+ network models. The images on the first row are CT images and the images on the second row are MR images. The areas marked in green represent the bone region prediction output of semantic segmentation network models.

In order to demonstrate the metric of our proposed CSN-ICP algorithm in point cloud registration, two other registration algorithms are adopted to perform the comparison, including ICP registration algorithm and RANSAC [29] registration algorithm. The sequential CT images possess two types of slices with 1.25 mm and 5 mm spacing and the difference of slice spacing is also considered in registration comparison experiments. The RMSE used for evaluate the point cloud registration is shown in Fig.6(a). The evaluation index $D_{MR}$ in orange line and $D_{CT}$ in green line for assessing the 2D CT and MR bone images is shown in Fig.6 (b). The triangles represent the registration between CT images of 1.25mm and MR images of 5mm. The circles represent the registration between CT images of 5mm and MR images of 5mm. The comparison results show that our proposed CSN-ICP can better reduce the point cloud RMSE in different slice spacing of CT images. The diagrams of CT and MR bone point cloud registration are shown in Fig.7. The point cloud registration on first row is employed with ICP algorithm, RANSAC algorithm and CSN-ICP algorithm. From the regions indicated by the arrow, the differences between different registration algorithm can be discerned clearly which also proves the validation when applying our method.

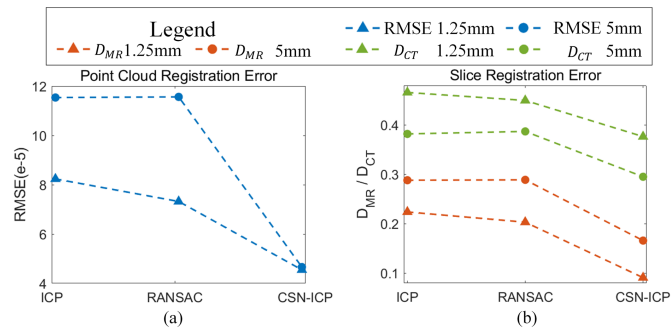

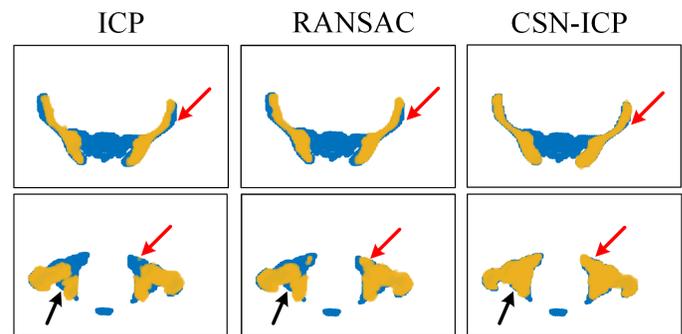

Fig.6 (a) Point cloud registration error indicator RMSE (unit: $10^{-5}$) in the ICP, RANSAC and CSN-ICP registration algorithms. (b) Slice registration error indicator $D_{MR}$ (orange line) and $D_{CT}$ (green line) in the ICP, RANSAC and CSN-ICP registration algorithms.

Arbitrary images can be extracted from registered 3D CT and MR point cloud. Through the joint coordinate system which combined the global Cartesian coordinate and local spherical coordinate, the description dimensions of each point in the point cloud are increased and the ability of error function to evaluate the matching degree of point cloud is improved. The slice images generated from CT and MR registered bone point cloud in the ICP, RANSAC and CSN-ICP registration algorithms are exhibited in Fig.8. The advantages of CSN-ICP registration algorithms can be shown from the area indicated by the arrow. The difference among registered images calculated by different registration algorithms shows that CSN-ICP registration algorithm can solve the problems of registration displacement based on partition registration related to bone characteristic. The registration results also show that the method proposed in this work can improve registration accuracy both on 3D point cloud and sequential images, which might help the medical practitioners in sparing pelvic bone marrow of cervical cancer radiotherapy.

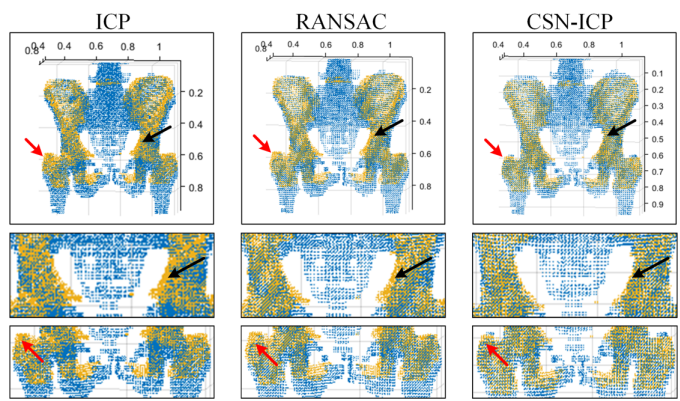

Fig.7. The diagrams of CT and MR bone point cloud registration respectively employed ICP, RANSAC and CSN-ICP algorithms on the first row. The images on the second row and the third row are zooming diagrams of the first row. The bone point clouds are displayed in normalized coordinate system.

Fig.8. The slice images generated from CT and MR registered bone point cloud with the ICP, RANSAC and CSN-ICP registration algorithms.

## IV. CONCLUSION

Bone marrow sparing using multimodal image is a criteria stage in radiotherapy for cervical cancer to enhance the living quality. A method based on point cloud construction from sequential multimodal images and CSN-ICP registration algorithm is proposed in this study with improvement of accuracy in both 3D point clouds and sequential images. The construction of point clouds can eliminate structural similarity difference in multimodal images. By introducing the joint coordinate system that combined global Cartesian coordinate system with local point cloud spherical coordinate system, the increasement of point descriptive dimension avoids the local optimal registration and improves the registration accuracy. The method proposed in this contribution might also provide a way to solve multimodal registration especially in sequential multimodal images in other clinical application.